\documentclass[aps,english,preprint,nofootinbib,preprintnumbers]{revtex4-2}
\usepackage{mathrsfs}
\usepackage{hyperref}
\hypersetup{
	colorlinks=true,
	linkcolor=blue,
	filecolor=magenta,      
	urlcolor=cyan,
	pdfpagemode=FullScreen,
}
\usepackage{graphicx}
\usepackage{amsmath, amssymb,bbm}
\usepackage{babel}
\usepackage{color}
\usepackage{slashed}
\usepackage[T1]{fontenc}
\usepackage{blindtext}
\usepackage{titlesec}
\usepackage{empheq}
\usepackage{multirow}
\def\beqn{\begin{eqnarray}}
	\def\eeqn{\end{eqnarray}}
\def\barr{\begin{array}}
	\def\earr{\end{array}}
\def\btab{\begin{tabular}}
	\def\etab{\end{tabular}}
\def\bite{\begin{itemize}}
	\def\eite{\end{itemize}}
\def\bcen{\begin{center}}
	\def\ecen{\end{center}}

\begin{document}
	
\title{Testing effective field theory with the most general neutron decay correlations}

\author{Chien-Yeah Seng$^{1,2}$}

\affiliation{$^{1}$Facility for Rare Isotope Beams, Michigan State University, East Lansing, MI 48824, USA}
\affiliation{$^{2}$Department of Physics, University of Washington,
Seattle, WA 98195-1560, USA}

\date{\today}

\begin{abstract}
		
Built on the seminal works by Jackson-Treiman-Wyld and Ebel-Feldman, we derive the most general free neutron differential decay rate where all massive particles (neutron, proton and electron) are polarized. This introduces 33 new correlations in addition to the 18 existing ones, which over-constrain the coupling constants in the low-energy effective field theory of charged weak interactions, and thus provides stringent tests of the validity of the theory framework itself. We classify the correlation coefficients in terms of their Standard Model limit and discrete symmetries, and study their expansion with respect to the new physics coupling strengths, supplemented by the experiment-independent $\mathcal{O}(\alpha)$ virtual electromagnetic radiative corrections.
		
\end{abstract}

\maketitle
\newpage
	
	
\section{Introduction}

The failure of the Standard Model (SM) to explain observed phenomena such as dark matter and the matter-antimatter asymmetry calls for the search of physics beyond the Standard Model (BSM). However, the fact that we have not yet observed any confirmed signal of new physics suggests that they either reside at very high energy scale, or may involve new light degrees of freedom (DOFs) with very weak couplings to the SM particle sector. For the former case, one may describe the remnant of new physics at low scales using the effective field theory (EFT) approach, where new heavy DOFs are integrated out in replacement of higher-order operators with the SM DOFs. At MeV scale, the effective theory for charged weak interactions relevant to neutron and nuclear beta decays is given by the Lee-Yang (LY) Hamiltonian~\cite{Lee:1956qn}, where the effective DOFs are nucleons, electrons, neutrinos and photons. By the renormalization group analysis, their coupling strengths can be related to the Wilson coefficients of the Low-Energy Effective Field Theory (LEFT)~\cite{Jenkins:2017jig} where quarks replace the nucleons as the active DOFs, and the Standard Model Effective Field Theory (SMEFT)~\cite{Buchmuller:1985jz,Grzadkowski:2010es} where heavy gauge bosons and quarks become active. 

The EFT description of BSM physics at low energies can break down if there are light new DOFs which cannot be integrated out. A direct consequence of such is that one may be unable (or find it unnatural) to consistently explain all experimental results using a finite number of EFT coupling constants at a given order. A recent example of this kind is the measurement of the axial-to-vector coupling constant $\lambda$ from the free neutron beta decay. This ratio has been extracted from the electron-neutrino correlation $a$ and the beta asymmetry parameter $A$; the best measurement of $a$ from aSPECT (with revised systematic errors and assuming the absence of Fierz term) returned $\lambda=-1.2668(27)$~\cite{Beck:2019xye,Beck:2023hnt}, while the best measurement of $A$ from PERKEO-III~\cite{Markisch:2018ndu} returned $\lambda=-1.27641(56)$, the two showed a 3.5$\sigma$ discrepancy. It was known that in the EFT framework, apart from the spectrum-modulation by the Fierz term $b$ that affects the experimentally-measured correlations:
\begin{equation}
	\mathcal{C}_\text{exp}(E_e)=\frac{\mathcal{C}(E_e)}{1+bm_e/E_e+\dots}~,
\end{equation}
the BSM corrections to both $a$ and $A$ occur at the second order~\cite{Gardner:2013aya,Cirigliano:2013xha}, which makes it difficult to explain the $a-A$ discrepancy while avoiding constraints from high-energy experiments, e.g. from the Large Hadron Collider (LHC)~\cite{Cirigliano:2012ab,Cirigliano:2013xha,Cirigliano:2023nol}, as well as the respective upper limits on the Fierz term set by the individual correlations~\cite{Hickerson:2017fzz,Gonzalez-Alonso:2016jzm,Gonzalez-Alonso:2018omy,Saul:2019qnp,Beck:2023hnt}.  
It is therefore desirable to search for more evidences of the possible breakdown of EFT by checking its consistency in the prediction of different correlations in the same decay process. 

In 1957, Jackson, Treiman and Wyld (JTW) \cite{Jackson:1957zz} wrote down a general formula for the free neutron differential decay rate, including terms that contain up to two out of the three following vector quantities: the neutron polarization $\hat{s}_n$, the electron polarization $\vec{\sigma}$, and the neutrino momentum $\vec{p}_\nu$ deduced from the proton recoil; Ebel and Feldman (EF)~\cite{ebel1957further} extended their result to include terms that involve all three vectors. However, to the best of our knowledge, the most general decay rate formula that includes the polarizations of \textit{all} massive particles: neutron, electron and proton, is not yet written down, the practical reason is simply that the proton polarization $\hat{s}_p$ is extremely difficult to measure and was not being attempted in existing experiments. However, at least in principle, the polarization of the final nuclei could be detected. Given the importance of having additional ways of verifying new physics, it would be interesting to consider the possibility that such measurements could be carried out. The polarization of the proton after neutron decay may be detectable if feasible ways of accelerating the proton without losing its polarization existed, allowing for analyzing by scattering. The polarization of nuclei after beta decays may lead to detectable effects in the final nucleus. This paper does not pretend to have a proposal for doing such experiments, but aims to inspire considerations for practical experiments.

The inclusion of $\hat{s}_p$ results in a much richer structure of the neutron decay phenomenology, with a total of 51 correlation coefficients that over-constrain the coupling constants of the beta decay EFT and offer a powerful consistency test of the theory framework and probe new light DOFs. In this work we attempt on such a generalization where all massive particles are allowed to be polarized. We also perform a classification of all the resulting correlations according to their discrete symmetries and properties under the expansion of small BSM parameters, and discuss their respective roles in the search for new physics.

\section{Most general set of correlations}
	
Weinberg's famous ``theorem'' of EFT~\cite{Weinberg:1978kz} states that, the most general Lagrangian $\mathcal{L}$ will give rise to the most general S-matrix consistent with analyticity, perturbative unitarity, cluster decomposition and the assumed symmetry principles in $\mathcal{L}$. 
Assuming that new DOFs are heavy, they can be integrated out at low energies to obtain higher-dimensional operators constructed from SM DOFs. At the beta decay energy scale, the active DOFs are nucleons and leptons, and the most general effective interactions at lowest order are given by the LY Hamiltonian~\cite{Lee:1956qn}:
\begin{eqnarray}
H_\text{LY}&=&\bar{p}n\left[C_S\bar{e}\nu-C_S'\bar{e}\gamma_5\nu\right]+\bar{p}\gamma^\mu n\left[C_V\bar{e}\gamma_\mu\nu-C_V'\bar{e}\gamma_\mu\gamma_5\nu\right]+\frac{1}{2}\bar{p}\sigma^{\mu\nu}n\left[C_T\bar{e}\sigma_{\mu\nu}\nu-C_T'\bar{e}\sigma_{\mu\nu}\gamma_5\nu\right]\nonumber\\
&&-\bar{p}\gamma^\mu\gamma_5n\left[C_A\bar{e}\gamma_\mu\gamma_5\nu-C_A'\bar{e}\gamma_\mu\nu\right]+\bar{p}\gamma_5n\left[C_P\bar{e}\gamma_5\nu-C_P'\bar{e}\nu\right]+h.c.~.\label{eq:HLY}
\end{eqnarray}
The most general correlations are then simply obtained by computing the free neutron decay squared amplitude from Eq.\eqref{eq:HLY}. Here we  allow all neutron, proton and electron to be polarized; at this point we do not consider higher-order effects such as radiative and recoil corrections. 

The desired squared amplitude at is given by:
\begin{equation}
	|\mathcal{M}|^2=4m_n m_pE_e E_\nu\xi\left\{1+g_\text{JTW}+g_\text{EF}+g_{s_p}+g_{s_ps_n}\right\}~,\label{eq:M2}
\end{equation}
which contains four classes of correlations apart from the overall spectrum factor $\xi$ (which will also be called a ``correlation'' throughout this paper). The first class consists of 12 terms in the JTW paper~\cite{Jackson:1957zz}:
\begin{eqnarray}
	g_\text{JTW}&=&a\frac{\vec{p}_e\cdot\vec{p}_\nu}{E_eE_\nu}+b\frac{m_e}{E_e}+\hat{s}_n\cdot\left[A\frac{\vec{p}_e}{E_e}+B\frac{\vec{p}_\nu}{E_\nu}+D\frac{\vec{p}_e\times \vec{p}_\nu}{E_eE_\nu}\right]\nonumber\\
	&&+\vec{\sigma}\cdot\left[G\frac{\vec{p}_e}{E_e}+H\frac{\vec{p}_\nu}{E_\nu}+K\frac{\vec{p}_e}{E_e+m_e}\frac{\vec{p}_e\cdot\vec{p}_\nu}{E_eE_\nu}+L\frac{\vec{p}_e\times\vec{p}_\nu}{E_eE_\nu}\right]\nonumber\\
	&&+\vec{\sigma}\cdot\left[N\hat{s}_n+Q\frac{\vec{p}_e}{E_e+m_e}\hat{s}_n\cdot\frac{\vec{p}_e}{E_e}+R\hat{s}_n\times\frac{\vec{p}_e}{E_e}\right]~.
\end{eqnarray}
The second class contains 5 correlations defined in the EF paper~\cite{ebel1957further}:
\begin{eqnarray}
	g_\text{EF}&=&S\hat{s}_n\cdot\vec{\sigma}\frac{\vec{p}_e\cdot\vec{p}_\nu}{E_e E_\nu}+T\frac{\vec{\sigma}\cdot\vec{p}_e}{E_e}\hat{s}_n\cdot\frac{\vec{p}_\nu}{E_\nu}+U\frac{\vec{\sigma}\cdot\vec{p}_\nu}{E_\nu}\hat{s}_n\cdot\frac{\vec{p}_e}{E_e}\nonumber\\
	&&+V\hat{s}_n\cdot\left(\frac{\vec{\sigma}\times\vec{p}_\nu}{E_\nu}\right)+W\frac{\vec{\sigma}\cdot\vec{p}_e}{E_e+m_e}\hat{s}_n\cdot\left(\frac{\vec{p}_e\times\vec{p}_\nu}{E_eE_\nu}\right)~.\label{eq:gEF}
\end{eqnarray}	
In the derivation of Eq.\eqref{eq:gEF}, we have used the following useful identity:
\begin{equation}
	\vec{a}\cdot\vec{b}~\vec{c}\cdot(\vec{d}\times\vec{e})-	\vec{a}\cdot\vec{c}~\vec{b}\cdot(\vec{d}\times\vec{e})+\vec{a}\cdot\vec{d}~\vec{b}\cdot(\vec{c}\times\vec{e})-\vec{a}\cdot\vec{e}~\vec{b}\cdot(\vec{c}\times\vec{d})=0
\end{equation}
to reduce the number of independent structures. Notice also that there are four more terms in Refs.\cite{Jackson:1957zz,ebel1957further}: $c$, $I$, $K'$ and $M$, that appear in a generic allowed beta decay but vanish in the case of the neutron decay where $J_i=J_f=1/2$. 

In this paper we introduce two more classes of correlations. The first class depends linearly on the proton unit polarization $\hat{s}_p$ but is independent of $\hat{s}_n$. Their structures can be directly inferred from the terms in $g_\text{JTW}$ and $g_\text{EF}$ by replacing $\hat{s}_n\rightarrow\hat{s}_p$. We find 11 new correlations:
\begin{eqnarray}
	g_{s_p}&=&\hat{s}_p\cdot\left[\tilde{A}\frac{\vec{p}_e}{E_e}+\tilde{B}\frac{\vec{p}_\nu}{E_\nu}+\tilde{D}\frac{\vec{p}_e\times\vec{p}_\nu}{E_eE_\nu}\right]+\vec{\sigma}\cdot\left[\tilde{N}\hat{s}_p+\tilde{Q}\frac{\vec{p}_e}{E_e+m_e}\hat{s}_p\cdot\frac{\vec{p}_e}{E_e}+\tilde{R}\hat{s}_p\times\frac{\vec{p}_e}{E_e}\right]\nonumber\\
	&&+\tilde{S}\hat{s}_p\cdot\vec{\sigma}\frac{\vec{p}_e\cdot\vec{p}_\nu}{E_e E_\nu}+\tilde{T}\frac{\vec{\sigma}\cdot\vec{p}_e}{E_e}\hat{s}_p\cdot\frac{\vec{p}_\nu}{E_\nu}+\tilde{U}\frac{\vec{\sigma}\cdot\vec{p}_\nu}{E_\nu}\hat{s}_p\cdot\frac{\vec{p}_e}{E_e}\nonumber\\
	&&+\tilde{V}\hat{s}_p\cdot\left(\frac{\vec{\sigma}\times\vec{p}_\nu}{E_\nu}\right)+\tilde{W}\frac{\vec{\sigma}\cdot\vec{p}_e}{E_e+m_e}\hat{s}_p\cdot\left(\frac{\vec{p}_e\times\vec{p}_\nu}{E_eE_\nu}\right)~.
\end{eqnarray}
Finally, there are structures that are linear simultaneously to $\hat{s}_n$ and $\hat{s}_p$, which have no analogy to the known structures and have to be worked out from scratch. There are 22 such terms:
\begin{eqnarray}
	g_{s_ps_n}&=&\hat{s}_p\cdot\hat{s}_n\left[X+\tilde{a}\frac{\vec{p}_e\cdot\vec{p}_\nu}{E_eE_\nu}+\tilde{b}\frac{m_e}{E_e}\right]+\hat{s}_p\cdot\hat{s}_n\vec{\sigma}\cdot\left[\tilde{G}\frac{\vec{p}_e}{E_e}+\tilde{H}\frac{\vec{p}_\nu}{E_\nu}+\tilde{K}\frac{\vec{p}_e}{E_e+m_e}\frac{\vec{p}_e\cdot\vec{p}_\nu}{E_eE_\nu}\right.\nonumber\\
	&&\left.+\tilde{L}\frac{\vec{p}_e\times\vec{p}_\nu}{E_eE_\nu}\right]+Y\vec{\sigma}\cdot(\hat{s}_p\times\hat{s}_n)+\text{\k{A}}\vec{\sigma}\cdot\hat{s}_p\frac{\vec{p}_\nu}{E_\nu}\cdot\hat{s}_n+\text{\k{a}}\vec{\sigma}\cdot\hat{s}_n\frac{\vec{p}_\nu}{E_\nu}\cdot\hat{s}_p\nonumber\\
	&&+\text{\k{E}}\vec{\sigma}\cdot\hat{s}_p\frac{\vec{p}_e}{E_e}\cdot\hat{s}_n+\text{\k{e}}\vec{\sigma}\cdot\hat{s}_n\frac{\vec{p}_e}{E_e}\cdot\hat{s}_p+\text{\L{}}\frac{\vec{p}_\nu}{E_\nu}\cdot(\hat{s}_p\times\hat{s}_n)+\text{\l{}}\frac{\vec{p}_e}{E_e}\cdot(\hat{s}_p\times\hat{s}_n)\nonumber\\
	&&+\text{\'{N}}\frac{\vec{p}_\nu}{E_\nu}\cdot\hat{s}_p\frac{\vec{p}_e}{E_e}\cdot\hat{s}_n+\text{\'{n}}\frac{\vec{p}_\nu}{E_\nu}\cdot\hat{s}_n\frac{\vec{p}_e}{E_e}\cdot\hat{s}_p+\text{\'{O}}\frac{\vec{p}_\nu}{E_\nu}\cdot\hat{s}_p\vec{\sigma}\cdot\left(\frac{\vec{p}_e}{E_e}\times\hat{s}_n\right)\nonumber\\
	&&+\text{\'{o}}\frac{\vec{p}_\nu}{E_\nu}\cdot\hat{s}_n\vec{\sigma}\cdot\left(\frac{\vec{p}_e}{E_e}\times\hat{s}_p\right)+\text{\'{S}}\frac{\vec{\sigma}\cdot\vec{p}_e}{E_e}\frac{\vec{p}_\nu}{E_\nu}\cdot(\hat{s}_p\times \hat{s}_n)+\text{\'{s}}\frac{\vec{\sigma}\cdot\vec{p}_e}{E_e+m_e}\frac{\vec{p}_e}{E_e}\cdot(\hat{s}_p\times \hat{s}_n)\nonumber\\
	&&+\text{\'{Z}}\frac{\vec{p}_\nu}{E_\nu}\cdot\hat{s}_p\frac{\vec{p}_e}{E_e}\cdot\hat{s}_n\frac{\vec{\sigma}\cdot\vec{p}_e}{E_e+m_e}+\text{\'{z}}\frac{\vec{p}_\nu}{E_\nu}\cdot\hat{s}_n\frac{\vec{p}_e}{E_e}\cdot\hat{s}_p\frac{\vec{\sigma}\cdot\vec{p}_e}{E_e+m_e}~.
\end{eqnarray} 
We label some of these new correlations with \textit{alfabet polski} as we are running out of Latin alphabets. 

\begin{table}
	\centering{}%
	\begin{tabular}{|c|c|c|}
		\hline 
		& P-even & P-odd\tabularnewline
		\hline 
		\multirow{3}{*}{T-even} & $\xi,a,b,N,Q,S,T,U,$  & $A,B,G,H,K,$\tabularnewline
		& $\tilde{N},\tilde{Q},\tilde{S},\tilde{T},\tilde{U},$ & $\tilde{A},\tilde{B},\tilde{G},\tilde{H},\tilde{K},$\tabularnewline
		& $X,\tilde{a},\tilde{b},\text{\'{N}},\text{\'{n}}$ & $\text{\k{A}},\text{\k{a}},\text{\k{E}},\text{\k{e}},\text{\'{Z}},\text{\'{z}}$\tabularnewline
		\hline 
		\multirow{3}{*}{T-odd} & $D,L,$ & $R,V,W,$\tabularnewline
		& $\tilde{D},\tilde{L},$ & $\tilde{R},\tilde{V},\tilde{W},$\tabularnewline
		& $Y,\text{\'{O}},\text{\'{o}},\text{\'{S}},\text{\'{s}}$ & $\text{\L{}},\text{\l{}}$\tabularnewline
		\hline 
	\end{tabular}\caption{\label{tab:symmetry}Discrete symmetries of the correlation coefficients.}
\end{table}

In Appendix~\ref{sec:zeroth} we provide the analytic formula for all 51 correlations induced by the pure LY Hamiltonian (i.e. without considering higher-order SM corrections); notice the absence of $C_P^{(\prime)}$ in the expressions due to the fact that $\bar{u}_p\gamma_5u_n\rightarrow 0$ in the non-recoil limit, despite possible enhancement from the pion pole~\cite{Gonzalez-Alonso:2013ura,Falkowski:2020pma}. It is also useful to study the discrete symmetries, i.e. parity (P) and time-reversal (T), of each correlation, which we summarize in Table~\ref{tab:symmetry}. Such information is important to relate precision neutron beta decay measurements to other experiments, both at high and low energies, that search for signals of new physics. For instance, T-odd interactions (which are also CP-odd given the CPT theorem) are important to explain the existing baryon-antibaryon asymmetry according to the Sakharov criteria \cite{Sakharov:1967dj}. Experimental measurement of the T-odd correlations in the neutron beta decay may combine with searches for permanent electric dipole moments~\cite{Chupp:2017rkp} and LHC experiments~\cite{Cirigliano:2012ab,El-Menoufi:2016cfo} to constrain the magnitude of CP-odd Wilson coefficients in SMEFT.

\section{BSM expansion}

The SM predicts a hierarchy of the different LY parameters, and for the purpose of new physics searches it is useful to perform an expansion of Eq.\eqref{eq:M2} from the SM limit. This has been considered in many literature for the few most commonly-studied correlation coefficients such as $\xi$, $a$, $b$, $A$, $B$ and $D$~\cite{Cirigliano:2012ab,Cirigliano:2013xha,Gardner:2013aya,Gonzalez-Alonso:2018omy,Falkowski:2020pma}, and here we shall generalize it to all 51 coefficients. 
	
To facilitate the discussion, we follow Ref.\cite{Falkowski:2020pma} and recombine the LY parameters as:
\begin{equation}
C_X=\frac{C_X^++C_X^-}{2}~,~C_X'=\frac{C_X^+-C_X^-}{2}~,~X=S,V,T,A,P~.
\end{equation}
With this, we can rewrite Eq.\eqref{eq:HLY} as:
\begin{eqnarray}
H_\text{LY}&=&\bar{p}n\left[C_S^+\bar{e}_R\nu_L+C_S^-\bar{e}_L\nu_R\right]+\bar{p}\gamma^\mu n\left[C_V^+\bar{e}_L\gamma_\mu\nu_L+C_V^-\bar{e}_R\gamma_\mu\nu_R\right]\nonumber\\
&&+\frac{1}{2}\bar{p}\sigma^{\mu\nu}n\left[C_T^+\bar{e}_R\sigma_{\mu\nu}\nu_L+C_T^-\bar{e}_L\sigma_{\mu\nu}\nu_R\right]+\bar{p}\gamma^\mu\gamma_5n\left[C_A^+\bar{e}_L\gamma_\mu\nu_L-C_A^-\bar{e}_R\gamma_\mu\nu_R\right]\nonumber\\
&&-\bar{p}\gamma_5n\left[C_P^+\bar{e}_R\nu_L-C_P^-\bar{e}_L\nu_R\right]+h.c.~,
\end{eqnarray}
i.e. the ``+''(``-'') coefficients are associated to left (right)-handed neutrinos, respectively. This representation is particularly useful since there are only two real parameters, namely $\mathfrak{Re}C_V^+$ and $\mathfrak{Re}C_A^+$, that have a zeroth-order SM contribution. We express them as:
\begin{equation}
\mathfrak{Re}C_V^+\equiv \sqrt{2}G_V~,~\mathfrak{Re}C_A^+\equiv \sqrt{2}G_V\lambda~,
\end{equation}
which defines the quantity $G_V$ and $\lambda$ (notice that $\lambda <0$ in this convention). Both quantities contain SM and BSM contributions: $G_V=(G_V)_\text{SM}+(G_V)_\text{BSM}$, $\lambda=(\lambda)_\text{SM}+(\lambda)_\text{BSM}$, with the former given by:
\begin{equation}
(G_V)_\text{SM}=G_FV_{ud}g_V~,~(\lambda)_\text{SM}=g_A/g_V~,
\end{equation}
where $G_F=1.1663788(6)\times 10^{-5}$~GeV$^{-2}$ is the Fermi constant measured from muon decay~\cite{MuLan:2012sih}, and $g_{V,A}$ are the vector and axial coupling constants in the neutron decay. Following common practice, we define $g_{V,A}$ to include the SM ``inner'' radiative corrections~\cite{Sirlin:1967zza,Gorchtein:2023srs}:
\begin{equation}
	g_{V,A}^2=\mathring{g}_{V,A}^2\left\{1+\Delta_{R}^{V,A}\right\}~,
\end{equation}
where the ``bare'' vector and axial coupling constants $\mathring{g}_{V,A}$ are defined through the pure-QCD matrix element of the charged weak current:
\begin{equation}
	\langle p|\bar{u}\gamma^\mu(1-\gamma_5)d|n\rangle_\text{QCD}=\bar{u}_p\gamma^\mu(\mathring{g}_V+\mathring{g}_A\gamma_5)u_n+\text{recoil}~.
\end{equation} 	
In particular, $\mathring{g}_V=1$ due to the conserved vector current, up to very small isospin-symmetry-breaking corrections that be studied using lattice QCD~\cite{Seng:2023jby}. The recent progress in the theory prediction of the inner radiative corrections $\Delta_R^{V,A}$~\cite{Seng:2018yzq,Shiells:2020fqp,Czarnecki:2019mwq,Seng:2020wjq,Hayen:2020cxh,Cirigliano:2022hob} is one of the main driving forces behind the resurgence of interest on the precision test of SM in charged weak decays, but will not be further discussed in this paper. 

When performing the BSM expansion, it is customary to keep the quantities $G_V$ and $\lambda$ intact without separating the SM (zeroth- and higher-order) and BSM pieces, because it is always the fully-renormalized quantities that are experimentally measured\footnote{With that said, it is still possible to probe new physics with these observables. For instance, comparing the experimentally-measured $\lambda$ to the theory prediction of $(\lambda)_\text{SM}$ may set constraints on BSM-induced right-handed quark currents~\cite{Bhattacharya:2011qm,Cirigliano:2022hob}.}. Therefore, we assign the following power counting for the expansion:
\begin{equation}
G_V,\lambda\sim\mathcal{O}(1)~,~\mathfrak{Im}C_{V,A}^+,C_{V,A}^-,C_{S,T,P}^{\pm}\sim \mathcal{O}(\epsilon)~,
\end{equation}
where $\epsilon$ counts the power of BSM coupling. With this, we can expand the LY-induced correlation coefficients in Appendix~\ref{sec:zeroth} with increasing powers of $\epsilon$:  
\begin{equation}
	(\mathcal{C}\xi)_\text{LY}=(\mathcal{C}\xi)_0+(\mathcal{C}\xi)_1+(\mathcal{C}\xi)_2+\dots\label{eq:Cexpand}
\end{equation}
It is easy to see that, up to $\mathcal{O}(\epsilon)$ only the ``+'' coefficients survive, because the SM neutrinos are purely left-handed, so the ``-'' coefficients that involve right-handed neutrinos do not survive in the interference terms assuming that neutrinos are massless. Furthermore, explicit calculation shows that $\mathfrak{Im}C_V^+$ and $\mathfrak{Im}C_A^+$ always appear as the combination $\mathfrak{Im}C_{AV}^+\equiv \mathfrak{Im}\{C_A^+-\lambda C_V^+\}$ at this order. Therefore, up to $\mathcal{O}(\epsilon)$ we have access to 7 real parameters in the LY Hamiltonian: $G_V$, $\lambda$, $\mathfrak{Im}C_{AV}^+$, $\mathfrak{Re}C_S^+$, $\mathfrak{Im}C_S^+$, $\mathfrak{Re}C_T^+$ and $\mathfrak{Im}C_T^+$. 

For numerical precision, it is necessary to include higher-order SM corrections on top of the LY interaction. Again, this has been discussed extensively in literature for the few most commonly-studied correlations, and here we take the first step to generalize it to all 51 terms. To $10^{-4}$ precision, the relevant higher-order effects are: (1) the recoil corrections, and (2) the electroweak radiative corrections. The recoil corrections can arise kinematically (e.g. from the nucleon kinetic energy) or dynamically (e.g. from the $1/m_N$-suppressed nucleon form factors)~\cite{Holstein:1974zf}, and it would be more convenient to treat all of them coherently under a unified framework such as the heavy-baryon chiral perturbation theory~\cite{Jenkins:1990jv}, which we defer to a future work. Meanwhile, the electroweak radiative corrections are of three types:
\begin{itemize}
	\item The long-distance Coulomb interaction between the outgoing proton and electron which gives rise to the Fermi function~\cite{Fermi:1934hr}
	\begin{eqnarray}
		F(E_e)=1+\frac{\pi\alpha }{\beta}+\mathcal{O}(\alpha^2)
	\end{eqnarray}
	(where $\beta=p_e/E_e$) as a multiplicative factor to the squared amplitude, and other ``Coulomb corrections'' to some correlation coefficients that can mimic T-odd effects~\cite{Jackson:1957auh,ebel1957further}. Notice that the Fermi function was traditionally obtained by solving the Dirac equation of the electron under a Coulomb potential, but recently there is a substantial effort to reformulate it in terms of EFT~\cite{Cirigliano:2023fnz,Hill:2023acw,Hill:2023bfh}.
	\item The ``inner'' electroweak radiative corrections $\Delta_R^{V,A}$ that renormalize the neutron vector and axial coupling constants.
	\item The remaining ``outer'' radiative corrections that modify the correlation coefficients. 
 \end{itemize}  
The first two are purely virtual (i.e. loop-induced) corrections, while the outer radiative corrections are a combination of virtual and real (i.e. bremsstrahlung, $n\rightarrow pe\nu\gamma$) effects. 

Unlike virtual corrections that do not alter the external kinematics, the bremsstrahlung contributions rely heavily on the actual experimental setup: For instance, the emitted photon (or even neutrino) can either be detected or undetected, and the various experimental cuts on the momenta of detected particles will affect the integration region of the four-body phase space. Early works by Sirlin~\cite{Sirlin:1967zza}, Shann~\cite{Shann:1971fz} and Garcia-Maya~\cite{Garcia:1978bq} expressed the radiatively-corrected differential decay rate in terms of the electron and the ``true'' neutrino momenta $\vec{p}_e$ and $\vec{p}_\nu$, which implicitly assumed that the neutrino is detected. It was later pointed out~\cite{Toth:1984ei} that, the na\"{\i}ve application of such a formula to most of the existing neutron decay experiments that do not directly detect the outcoming neutrino can lead to an $\mathcal{O}(\alpha/\pi)$ error in the extraction of the $\vec{p}_\nu$-dependent correlation coefficients, due to the breakdown of the tree-level momentum conservation $\vec{p}_\nu=\vec{p}_n-\vec{p}_p-\vec{p}_e$ in the presence of the extra photon. This could be remedied by adopting a modified analysis of the bremsstrahlung corrections that fully integrates out the neutrino and photon momenta in the phase space integral, leaving instead $\vec{p}_p$ and $\vec{p}_e$ as free variables~\cite{Gluck:2022ogz,Seng:2023ynd}. Such a prescription is, however, inappropriate for experiments that explicitly detect the emitted photon (above certain momentum).

For the sake of generality, we decide to include in this work only the ``experiment-independent'' part of the radiative corrections, i.e. the virtual corrections, to $\mathcal{O}(\alpha)$. The resulting differential decay rate is thus incomplete (and in fact infrared-divergent), and must be supplemented by the corresponding bremsstrahlung correction calculation tailored to meet the actual setup of the specific experiment to be analyzed. The virtually-corrected differential decay rate takes the following form:
\begin{equation}
	\frac{d\Gamma}{dE_ed\Omega_ed\Omega_\nu}\approx \frac{p_e E_\nu}{512\pi^5m_nm_p}F(E_e)\left(1+\frac{\alpha}{2\pi}\delta_v^U\right)|\mathcal{M}|^2~,
\end{equation}	
where $F(E_e)$ is the Fermi function, and $\delta_v^U$ is the ``universal'' part of the virtual outer radiative corrections\footnote{There is a freedom to shift $\delta_v^U$ by an arbitrary additive constant, which is compensated by a corresponding shift in $\Delta_R^{V,A}$ simultaneously. Our choice of $\delta_v^U$ in Eq.\eqref{eq:deltavU} is consistent to the ``standard'' definition of $\Delta_R^V$ from which numerical values are quoted, see. e.g. \cite{Gorchtein:2023srs} and references therein.}:
\begin{equation}
	\delta_v^U=\frac{1}{2}\ln\frac{m_p^2}{m_e^2}-\frac{11}{4}+\ln\frac{m_e^2m_p^2}{m_\gamma^4}+\frac{2}{\beta}\tanh^{-1}\beta\left[\ln\frac{m_\gamma^2}{m_e^2}-\tanh^{-1}\beta\right]-\frac{2}{\beta}\text{Li}_2\left(\frac{2\beta}{1+\beta}\right)~,\label{eq:deltavU}
\end{equation}
where $m_\gamma$ is a fictitious photon mass to regularize the infrared divergence. The squared amplitude $|\mathcal{M}|^2$ takes the form of Eq.\eqref{eq:M2}, with the correlation coefficients expanded in increasing powers of $\epsilon$ (see Eq.\eqref{eq:Cexpand}) and supplemented by correlation-dependent virtual outer radiative corrections:
\begin{equation}
	\mathcal{C}\xi=\left(1+\frac{\alpha}{2\pi}\delta_{v\text{I}}^\mathcal{C}\right)(\mathcal{C}\xi)_0+G_V^2\frac{\alpha}{2\pi}\delta_{v\text{II}}^\mathcal{C}+(\mathcal{C}\xi)_1+\dots
\end{equation}
In particular, $\delta_{v\text{II}}^\text{C}$ represents the ``Coulomb correction'' that mimics T-odd correlations. We do not include the recoil-suppressed radiative corrections~\cite{Ivanov:2019rkp,Ivanov:2021bae} that can be relevant to certain observables; for example, it was known that the leading SM contribution to the $D$-coefficient scales as $\mathcal{O}(\alpha E_e/m_N)\sim 10^{-5}$~\cite{Callan:1967zz}.

\begin{table}
	\begin{centering}
		\begin{tabular}{|c|c|c|c|c|}
			\hline 
			$\mathcal{C}$ & $(\mathcal{C}\xi)_{0}$ & $\delta_{v\text{I}}^{\mathcal{C}}$ & $\delta_{v\text{II}}^{\mathcal{C}}$ & $(\mathcal{C}\xi)_{1}$\tabularnewline
			\hline 
			\hline 
			1 & $G_{V}^{2}(1+3\lambda^{2})$ & $2\beta\tanh^{-1}\beta$ & 0 & 0\tabularnewline
			\hline 
			$a$ & $-G_{V}^{2}(\lambda^{2}-1)$ & $\frac{2}{\beta}\tanh^{-1}\beta$ & 0 & 0\tabularnewline
			\hline 
			$b$ & 0 & 0 & 0 & $\sqrt{2}G_{V}\mathfrak{Re}\left\{ C_{S}^{+}+3\lambda C_{T}^{+}\right\} $\tabularnewline
			\hline 
			$A$ & $-2G_{V}^{2}\lambda(\lambda+1)$ & $\frac{2}{\beta}\tanh^{-1}\beta$ & 0 & 0\tabularnewline
			\hline 
			$B$ & $2G_{V}^{2}\lambda(\lambda-1)$ & $2\beta\tanh^{-1}\beta$ & 0 & $-\sqrt{2}G_{V}\frac{m_{e}}{E_{e}}\mathfrak{Re}\left\{ \lambda C_{S}^{+}+(1-2\lambda)C_{T}^{+}\right\} $\tabularnewline
			\hline 
			$D$ & 0 & 0 & 0 & $\sqrt{2}G_{V}\mathfrak{Im}C_{AV}^{+}$\tabularnewline
			\hline 
			$G$ & $-G_{V}^{2}(1+3\lambda^{2})$ & $\frac{2}{\beta}\tanh^{-1}\beta$ & 0 & 0\tabularnewline
			\hline 
			$H$ & $G_{V}^{2}\frac{m_{e}}{E_{e}}(\lambda^{2}-1)$ & 0 & 0 & $-\sqrt{2}G_{V}\mathfrak{Re}\left\{ C_{S}^{+}-\lambda C_{T}^{+}\right\} $\tabularnewline
			\hline 
			$K$ & $G_{V}^{2}(\lambda^{2}-1)$ & $\frac{2}{\beta}\frac{E_{e}+m_{e}}{E_{e}}\tanh^{-1}\beta$ & 0 & $\sqrt{2}G_{V}\mathfrak{Re}\left\{ C_{S}^{+}-\lambda C_{T}^{+}\right\} $\tabularnewline
			\hline 
			$L$ & 0 & 0 & $-\frac{2\pi(\lambda^{2}-1)m_{e}}{p_{e}}$ & $\sqrt{2}G_{V}\mathfrak{Im}\left\{ C_{S}^{+}-\lambda C_{T}^{+}\right\} $\tabularnewline
			\hline 
			$N$ & $2G_{V}^{2}\frac{m_{e}}{E_{e}}\lambda(\lambda+1)$ & 0 & 0 & $\sqrt{2}G_{V}\mathfrak{Re}\left\{ \lambda C_{S}^{+}+(2\lambda+1)C_{T}^{+}\right\} $\tabularnewline
			\hline 
			$Q$ & $2G_{V}^{2}\lambda(\lambda+1)$ & $\frac{2}{\beta}\frac{E_{e}+m_{e}}{E_{e}}\tanh^{-1}\beta$ & 0 & $-\sqrt{2}G_{V}\mathfrak{Re}\left\{ \lambda C_{S}^{+}+(2\lambda+1)C_{T}^{+}\right\} $\tabularnewline
			\hline 
			$R$ & 0 & 0 & $\frac{4\pi\lambda(\lambda+1)m_{e}}{p_{e}}$ & $\sqrt{2}G_{V}\mathfrak{Im}\left\{ \lambda C_{S}^{+}+(2\lambda+1)C_{T}^{+}\right\} $\tabularnewline
			\hline 
		\end{tabular}
		\par\end{centering}
	\caption{\label{tab:gJTW}Relevant quantities in the expansion of $\xi$ and $g_{\text{JTW}}$.}
	
\end{table}

\begin{table}
	\begin{centering}
		\begin{tabular}{|c|c|c|c|c|}
			\hline 
			$\mathcal{C}$ & $(\mathcal{C}\xi)_{0}$ & $\delta_{v\text{I}}^{\mathcal{C}}$ & $\delta_{v\text{II}}^{\mathcal{C}}$ & $(\mathcal{C}\xi)_{1}$\tabularnewline
			\hline 
			\hline 
			$S$ & 0 & 0 & 0 & $\sqrt{2}G_{V}\mathfrak{Re}\left\{ C_{T}^{+}-\lambda C_{S}^{+}\right\} $\tabularnewline
			\hline 
			$T$ & $-2G_{V}^{2}\lambda(\lambda-1)$ & $\frac{2}{\beta}\tanh^{-1}\beta$ & 0 & 0\tabularnewline
			\hline 
			$U$ & 0 & 0 & 0 & $-\sqrt{2}G_{V}\mathfrak{Re}\left\{ C_{T}^{+}-\lambda C_{S}^{+}\right\} $\tabularnewline
			\hline 
			$V$ & 0 & 0 & 0 & $-\sqrt{2}G_{V}\mathfrak{Im}\left\{ C_{T}^{+}-\lambda C_{S}^{+}+\frac{m_{e}}{E_{e}}C_{AV}^{+}\right\} $\tabularnewline
			\hline 
			$W$ & 0 & 0 & 0 & $-\sqrt{2}G_{V}\mathfrak{Im}\left\{ C_{AV}^{+}+\lambda C_{S}^{+}-C_{T}^{+}\right\} $\tabularnewline
			\hline 
		\end{tabular}
		\par\end{centering}
	\caption{\label{tab:gEF}Relevant quantities in the expansion of $g_{\text{EF}}$.}
	
\end{table}

\begin{table}
	\begin{centering}
		\begin{tabular}{|c|c|c|c|c|}
			\hline 
			$\mathcal{C}$ & $(\mathcal{C}\xi)_{0}$ & $\delta_{v\text{I}}^{\mathcal{C}}$ & $\delta_{v\text{II}}^{\mathcal{C}}$ & $(\mathcal{C}\xi)_{1}$\tabularnewline
			\hline 
			\hline 
			$\tilde{A}$ & $2G_{V}^{2}\lambda(\lambda-1)$ & $\frac{2}{\beta}\tanh^{-1}\beta$ & 0 & 0\tabularnewline
			\hline 
			$\tilde{B}$ & $-2G_{V}^{2}\lambda(\lambda+1)$ & $2\beta\tanh^{-1}\beta$ & 0 & $-\sqrt{2}G_{V}\frac{m_{e}}{E_{e}}\mathfrak{Re}\left\{ \lambda C_{S}^{+}+(2\lambda+1)C_{T}^{+}\right\} $\tabularnewline
			\hline 
			$\tilde{D}$ & 0 & 0 & 0 & $\sqrt{2}G_{V}\mathfrak{Im}C_{AV}^{+}$\tabularnewline
			\hline 
			$\tilde{N}$ & $-2G_{V}^{2}\frac{m_{e}}{E_{e}}\lambda(\lambda-1)$ & 0 & 0 & $\sqrt{2}G_{V}\mathfrak{Re}\left\{ \lambda C_{S}^{+}+(1-2\lambda)C_{T}^{+}\right\} $\tabularnewline
			\hline 
			$\tilde{Q}$ & $-2G_{V}^{2}\lambda(\lambda-1)$ & $\frac{2}{\beta}\frac{E_{e}+m_{e}}{E_{e}}\tanh^{-1}\beta$ & 0 & $-\sqrt{2}G_{V}\mathfrak{Re}\left\{ \lambda C_{S}^{+}+(1-2\lambda)C_{T}^{+}\right\} $\tabularnewline
			\hline 
			$\tilde{R}$ & 0 & 0 & $-\frac{4\pi\lambda(\lambda-1)m_{e}}{p_{e}}$ & $\sqrt{2}G_{V}\mathfrak{Im}\left\{ \lambda C_{S}^{+}+(1-2\lambda)C_{T}^{+}\right\} $\tabularnewline
			\hline 
			$\tilde{S}$ & 0 & 0 & 0 & $\sqrt{2}G_{V}\mathfrak{Re}\left\{ C_{T}^{+}-\lambda C_{S}^{+}\right\} $\tabularnewline
			\hline 
			$\tilde{T}$ & $2G_{V}^{2}\lambda(\lambda+1)$ & $\frac{2}{\beta}\tanh^{-1}\beta$ & 0 & 0\tabularnewline
			\hline 
			$\tilde{U}$ & 0 & 0 & 0 & $-\sqrt{2}G_{V}\mathfrak{Re}\left\{ C_{T}^{+}-\lambda C_{S}^{+}\right\} $\tabularnewline
			\hline 
			$\tilde{V}$ & 0 & 0 & 0 & $-\sqrt{2}G_{V}\mathfrak{Im}\left\{ C_{T}^{+}-\lambda C_{S}^{+}+\frac{m_{e}}{E_{e}}C_{AV}^{+}\right\} $\tabularnewline
			\hline 
			$\tilde{W}$ & 0 & 0 & 0 & $-\sqrt{2}G_{V}\mathfrak{Im}\left\{ C_{AV}^{+}+\lambda C_{S}^{+}-C_{T}^{+}\right\} $\tabularnewline
			\hline 
		\end{tabular}
		\par\end{centering}
	\caption{\label{tab:gsp}Relevant quantities in the expansion of $g_{s_{p}}$.}
\end{table}

\begin{table}
	\begin{centering}
		\begin{tabular}{|c|c|c|c|c|}
			\hline 
			$\mathcal{C}$ & $(\mathcal{C}\xi)_{0}$ & $\delta_{v\text{I}}^{\mathcal{C}}$ & $\delta_{v\text{II}}^{\mathcal{C}}$ & $(\mathcal{C}\xi)_{1}$\tabularnewline
			\hline 
			\hline 
			$X$ & $-G_{V}^{2}(\lambda^{2}-1)$ & $2\beta\tanh^{-1}\beta$ & 0 & 0\tabularnewline
			\hline 
			$\tilde{a}$ & $-G_{V}^{2}(\lambda^{2}-1)$ & $\frac{2}{\beta}\tanh^{-1}\beta$ & 0 & 0\tabularnewline
			\hline 
			$\tilde{b}$ & 0 & 0 & 0 & $\sqrt{2}G_{V}\mathfrak{Re}\left\{ C_{S}^{+}-\lambda C_{T}^{+}\right\} $\tabularnewline
			\hline 
			$\tilde{G}$ & $G_{V}^{2}(\lambda^{2}-1)$ & $\frac{2}{\beta}\tanh^{-1}\beta$ & 0 & 0\tabularnewline
			\hline 
			$\tilde{H}$ & $G_{V}^{2}\frac{m_{e}}{E_{e}}(\lambda^{2}-1)$ & 0 & 0 & $-\sqrt{2}G_{V}\mathfrak{Re}\left\{ C_{S}^{+}-\lambda C_{T}^{+}\right\} $\tabularnewline
			\hline 
			$\tilde{K}$ & $G_{V}^{2}(\lambda^{2}-1)$ & $\frac{2}{\beta}\frac{E_{e}+m_{e}}{E_{e}}\tanh^{-1}\beta$ & 0 & $\sqrt{2}G_{V}\mathfrak{Re}\left\{ C_{S}^{+}-\lambda C_{T}^{+}\right\} $\tabularnewline
			\hline 
			$\tilde{L}$ & 0 & 0 & $-\frac{2\pi(\lambda^{2}-1)m_{e}}{p_{e}}$ & $\sqrt{2}G_{V}\mathfrak{Im}\left\{ C_{S}^{+}-\lambda C_{T}^{+}\right\} $\tabularnewline
			\hline 
			$Y$ & 0 & 0 & 0 & $\sqrt{2}G_{V}\mathfrak{Im}\left\{ C_{T}^{+}-\lambda C_{S}^{+}+\frac{m_{e}}{E_{e}}C_{AV}^{+}\right\} $\tabularnewline
			\hline 
			\k{A}& $-2G_{V}^{2}\frac{m_{e}}{E_{e}}\lambda(\lambda-1)$ & 0 & 0 & $\sqrt{2}G_{V}\mathfrak{Re}\left\{ \lambda C_{S}^{+}+(1-2\lambda)C_{T}^{+}\right\} $\tabularnewline
			\hline 
			\k{a} & $-2G_{V}^{2}\frac{m_{e}}{E_{e}}\lambda(\lambda+1)$ & 0 & 0 & $-\sqrt{2}G_{V}\mathfrak{Re}\left\{ \lambda C_{S}^{+}+(2\lambda+1)C_{T}^{+}\right\} $\tabularnewline
			\hline 
			\k{E} & 0 & 0 & 0 & $\sqrt{2}G_{V}\mathfrak{Re}\left\{ C_{T}^{+}-\lambda C_{S}^{+}\right\} $\tabularnewline
			\hline 
			\k{e} & 0 & 0 & 0 & $-\sqrt{2}G_{V}\mathfrak{Re}\left\{ C_{T}^{+}-\lambda C_{S}^{+}\right\} $\tabularnewline
			\hline 
			\L{} & 0 & 0 & 0 & $-\sqrt{2}G_{V}\mathfrak{Im}\left\{ C_{AV}^{+}+\frac{m_{e}}{E_{e}}\left(C_{T}^{+}-\lambda C_{S}^{+}\right)\right\} $\tabularnewline
			\hline 
			\l{} & 0 & 0 & 0 & $-\sqrt{2}G_{V}\mathfrak{Im}C_{AV}^{+}$\tabularnewline
			\hline 
			\'{N} & $2G_{V}^{2}\lambda(\lambda+1)$ & $\frac{2}{\beta}\tanh^{-1}\beta$ & 0 & 0\tabularnewline
			\hline 
			\'{n} & $2G_{V}^{2}\lambda(\lambda-1)$ & $\frac{2}{\beta}\tanh^{-1}\beta$ & 0 & 0\tabularnewline
			\hline 
			\'{O} & 0 & 0 & $\frac{4\pi\lambda(\lambda+1)m_{e}}{p_{e}}$ & $\sqrt{2}G_{V}\mathfrak{Im}\left\{ \lambda C_{S}^{+}+(2\lambda+1)C_{T}^{+}\right\} $\tabularnewline
			\hline 
			\'{o} & 0 & 0 & $\frac{4\pi\lambda(\lambda-1)m_{e}}{p_{e}}$ & $-\sqrt{2}G_{V}\mathfrak{Im}\left\{ \lambda C_{S}^{+}+(1-2\lambda)C_{T}^{+}\right\} $\tabularnewline
			\hline 
			\'{S} & 0 & 0 & 0 & $\sqrt{2}G_{V}\mathfrak{Im}C_{AV}^{+}$\tabularnewline
			\hline 
			\'{s} & 0 & 0 & 0 & $\sqrt{2}G_{V}\mathfrak{Im}\left\{ C_{AV}^{+}+\lambda C_{S}^{+}-C_{T}^{+}\right\} $\tabularnewline
			\hline 
			\'{Z} & $-2G_{V}^{2}\lambda(\lambda+1)$ & $\frac{2}{\beta}\frac{E_{e}+m_{e}}{E_{e}}\tanh^{-1}\beta$ & 0 & $\sqrt{2}G_{V}\mathfrak{Re}\left\{ \lambda C_{S}^{+}+(2\lambda+1)C_{T}^{+}\right\} $\tabularnewline
			\hline 
			\'{z} & $-2G_{V}^{2}\lambda(\lambda-1)$ & $\frac{2}{\beta}\frac{E_{e}+m_{e}}{E_{e}}\tanh^{-1}\beta$ & 0 & $-\sqrt{2}G_{V}\mathfrak{Re}\left\{ \lambda C_{S}^{+}+(1-2\lambda)C_{T}^{+}\right\} $\tabularnewline
			\hline 
		\end{tabular}
		\par\end{centering}
	\caption{\label{tab:gspsn}Relevant quantities in the expansion of $g_{s_{p}s_{n}}$.}
\end{table}

\begin{table}
	\centering{}%
	\begin{tabular}{|c|c|c|c|}
		\hline 
		\multicolumn{2}{|c|}{Class A: Starts at $\mathcal{O}$(1)} & \multicolumn{2}{c|}{Class B: Starts at $\mathcal{O}(\epsilon)$}\tabularnewline
		\hline 
		I: Non-zero $\mathcal{O}(\epsilon)$ & II: Zero $\mathcal{O}(\epsilon)$ & I: T-even & II: T-odd\tabularnewline
		\hline 
		$B,H,K,N,Q,$ & $\xi,a,A,G,T,$ & $b,S,U,$ & $D,L,R,V,W,$\tabularnewline
		$\tilde{B},\tilde{N},\tilde{Q},\tilde{H},\tilde{K},$ & $\tilde{A},\tilde{T},X,\tilde{a},\tilde{G},$ & $\tilde{S},\tilde{U},\tilde{b},$ & $\tilde{D},\tilde{R},\tilde{V},\tilde{W},\tilde{L},$\tabularnewline
		$\text{\k{A}},\text{\k{a}},\text{\'{Z}},\text{\'{z}}$ & $\text{\'{N}},\text{\'{n}}$ & $\text{\k{E}},\text{\k{e}}$ & $Y,\text{\L{}},\text{\l{}},\text{\'{O}},\text{\'{o}},\text{\'{S}},\text{\'{s}}$\tabularnewline
		\hline 
	\end{tabular}\caption{\label{tab:classify}Classification of the correlation coefficients under the BSM expansion.}
\end{table}

In Table~\ref{tab:gJTW}-\ref{tab:gspsn}, we summarize all the relevant quantities in expansion of the 51 terms up to $\mathcal{O}(\epsilon)$: $(\mathcal{C}\xi)_0$, $\delta_{v\text{I}}^\mathcal{C}$, $\delta_{v\text{II}}^\mathcal{C}$, $(\mathcal{C}\xi)_1$. 

\section{SM, BSM and EFT}

For a more intuitive understanding, it is useful to categorize the correlations into two different classes (see Table~\ref{tab:classify}): Class-A elements which begin at $\mathcal{O}(1)$, and Class-B elements which begin at $\mathcal{O}(\epsilon)$; the former provide avenues for the precision extraction of of $G_V$ and $\lambda$ (and new physics parameters provided the SM background are well-understood), whereas the latter have a reduced SM background and are particularly useful to search for new physics. We further divide the Class-A elements into group I that receives $\mathcal{O}(\epsilon)$ corrections from the BSM-induced LY parameters, and group II that does not receive $\mathcal{O}(\epsilon)$ corrections. This is useful for the precision test of the EFT framework: For instance, if the extractions of $\lambda$ from two different Class-A.II correlations show a discrepancy, then its explanation within the EFT framework would involve BSM physics at the order $\mathcal{O}(\epsilon^2)$ and could be challenging (e.g. in order to comply with constraints from high-energy experiments); this could point towards the existence of light DOFs. We similarly divide the Class-B elements into group I that is T-even and group II that is T-odd (up to small SM Coulomb corrections). This is useful for tests of fundamental symmetries: For instance, the Class-B.II correlations are sensitive to CP-violating interactions needed for baryogenesis. 

\begin{table}
	\begin{centering}
		\begin{tabular}{|c|c|}
			\hline 
			& EFT-suppressed combinations\tabularnewline
			\hline 
			(JTW,EF)$\oplus$(JTW,EF) & $S+U$\tabularnewline
			\hline 
			\multirow{2}{*}{JTW$\oplus$new} & $a-\tilde{a}$, $a+\tilde{G}$, $A+\tilde{T}$, $A+\text{\'{N}}$, $D-\tilde{D}$,
			$D+\text{\l{}}$, $D-\text{\'{S}}$\tabularnewline
			& $H-\tilde{H}$, $K-\tilde{K}$, $L-\tilde{L}$, $N+\text{\k{a}}$, $Q+\text{\'{Z}}$, $R-\text{\'{O}}$ \tabularnewline
			\hline 
			\multirow{2}{*}{EF$\oplus$new} & $S-\tilde{S}$, $S+\tilde{U}$, $S-\text{\k{E}}$, $S+\text{\k{e}}$, $T+\tilde{A}$, $T+\text{\'{n}}$, $U+\tilde{S}$ \tabularnewline
			& $U-\tilde{U}$, $U+\text{\k{E}}$, $U-\text{\k{e}}$, $V-\tilde{V}$, $V+Y$, $W-\tilde{W}$,
			$W+\text{\'{s}}$\tabularnewline
			\hline 
		\end{tabular}
		\par\end{centering}
	\caption{\label{tab:zerosum}A list of sum between two correlation coefficients that give vanishing $\mathcal{O}(1)$, $\mathcal{O}(\epsilon)$ terms and the $\mathcal{O}(\alpha)$ virtual radiative corrections. We only display terms that involve up to one ``new'' correlation coming from $g_{s_p}$ or $g_{s_ps_n}$. }
	
\end{table}

An interesting application of the analysis above is to find possible sums of two correlation coefficients that have vanishing $\mathcal{O}(1)$, $\mathcal{O}(\epsilon)$ contributions and virtual radiative corrections; this means in the EFT framework such sums are extremely suppressed (both in SM and BSM), so a significant non-zero experimental value would be a strong evidence of the breakdown of EFT. When restricting to the ``old'' correlations, namely those from $g_\text{JTW}$ and $g_\text{EF}$, we only find one such sum: $S+U$, but with the inclusion of $g_{s_p}$ and $g_{s_p s_n}$ we find a lot more possibilities, see Table~\ref{tab:zerosum}. This again supports our assertion that the future measurement of the proton polarization will open new windows to test the EFT framework. 

We conclude this section by briefly discussing the $\mathcal{O}(\epsilon^2)$ corrections. They can be obtained by expanding the full expressions in Appendix~\ref{sec:zeroth} and are in general quite lengthy; here we show the results for a few most frequently-studied correlations:
\begin{eqnarray}
(\xi)_2&=&\frac{1}{2}\left\{|C_S^+|^2+|C_S^-|^2+|C_V^-|^2+3\left(|C_T^+|^2+|C_T^-|^2+|C_A^-|^2\right)+(\mathfrak{Im}C_V^+)^2+3(\mathfrak{Im}C_A^+)^2\right\}\nonumber\\
(a\xi)_2&=&\frac{1}{2}\left\{-|C_S^+|^2-|C_S^-|^2+|C_T^+|^2+|C_T^-|^2+|C_V^-|^2-|C_A^-|^2+(\mathfrak{Im}C_V^+)^2-(\mathfrak{Im}C_A^+)^2\right\}\nonumber\\
(A\xi)_2&=&|C_T^+|^2-|C_T^-|^2+|C_A^-|^2-(\mathfrak{Im}C_A^+)^2+\mathfrak{Re}C_S^+\mathfrak{Re}C_T^+-\mathfrak{Re}C_S^-\mathfrak{Re}C_T^-+\mathfrak{Re}C_V^-\mathfrak{Re}C_A^-\nonumber\\
&&+\mathfrak{Im}C_S^+\mathfrak{Im}C_T^+-\mathfrak{Im}C_S^-\mathfrak{Im}C_T^--\mathfrak{Im}C_V^+\mathfrak{Im}C_A^++\mathfrak{Im}C_V^-\mathfrak{Im}C_A^-~.
\end{eqnarray}

As an application, we consider the $a-A$ discrepancy from aSPECT and PERKEO-III. The two experiments had included an independent fit of the Fierz term, which returned $b=-0.0098(193)$~\cite{Beck:2023hnt} and $0.017(21)$~\cite{Saul:2019qnp} respectively, both consistent with zero; so we may assume $b=0$ in what follows for simplicity. We may try to explain the discrepancy between the $\lambda$ measured from $a$ and $A$ (after accounting for SM corrections), which we denote as $\lambda_a$ and $\lambda_A$ respectively, in terms of $\mathcal{O}(\epsilon^2)$ effects. This amounts to writing:
\begin{equation}
	\frac{1-\lambda_a^2}{1+3\lambda_a^2}=\frac{G_V^2(1-\lambda^2)+(a\xi)_2}{G_V^2(1+3\lambda^2)+(\xi)_2}~,~\frac{-2\lambda_A(\lambda_A+1)}{1+3\lambda_A^2}=\frac{-2G_V^2\lambda(\lambda+1)+(A\xi)_2}{G_V^2(1+3\lambda^2)+(\xi)_2}~,
\end{equation}
and expanding both expressions to $\mathcal{O}(\epsilon^2)$. That gives:
\begin{equation}
	\lambda_a-\lambda_A\approx \frac{1+3\lambda^2}{8G_V^2\lambda(1-\lambda)(1+3\lambda)}\left[(1+\lambda)^2(\xi)_2+4\lambda(A\xi)_2-(1-\lambda)(1+3\lambda)(a\xi)_2\right]~.\label{eq:lambdaaA}
\end{equation}
One can then analyze if such explanation is compatible with constraints on BSM Wilson coefficients from high-energy experiments. Plugging the experimental results from aSPECT and PERKEO-III to the left hand side returns $\lambda_a-\lambda_A=9.6(2.8)\times 10^{-3}$. Meanwhile, by dimensional analysis we assume that the BSM couplings scale as $1/\Lambda_\text{BSM}^2$ where $\Lambda_\text{BSM}$ is a new physics energy scale. So, up to order-one factors, the $a-A$ discrepancy would imply a relatively low new physics scale:
\begin{equation}
	\frac{1}{G_V^2\Lambda_\text{BSM}^4}\sim \lambda_a-\lambda_A\implies \Lambda_\text{BSM}\sim 1~\text{TeV}
\end{equation}
which is difficult to avoid LHC bounds. Similar analysis can be performed if more Class-A.II correlations are experimentally studied in the future. 

\section{Model example}

\begin{figure}
	\begin{centering}
		\includegraphics[scale=0.15]{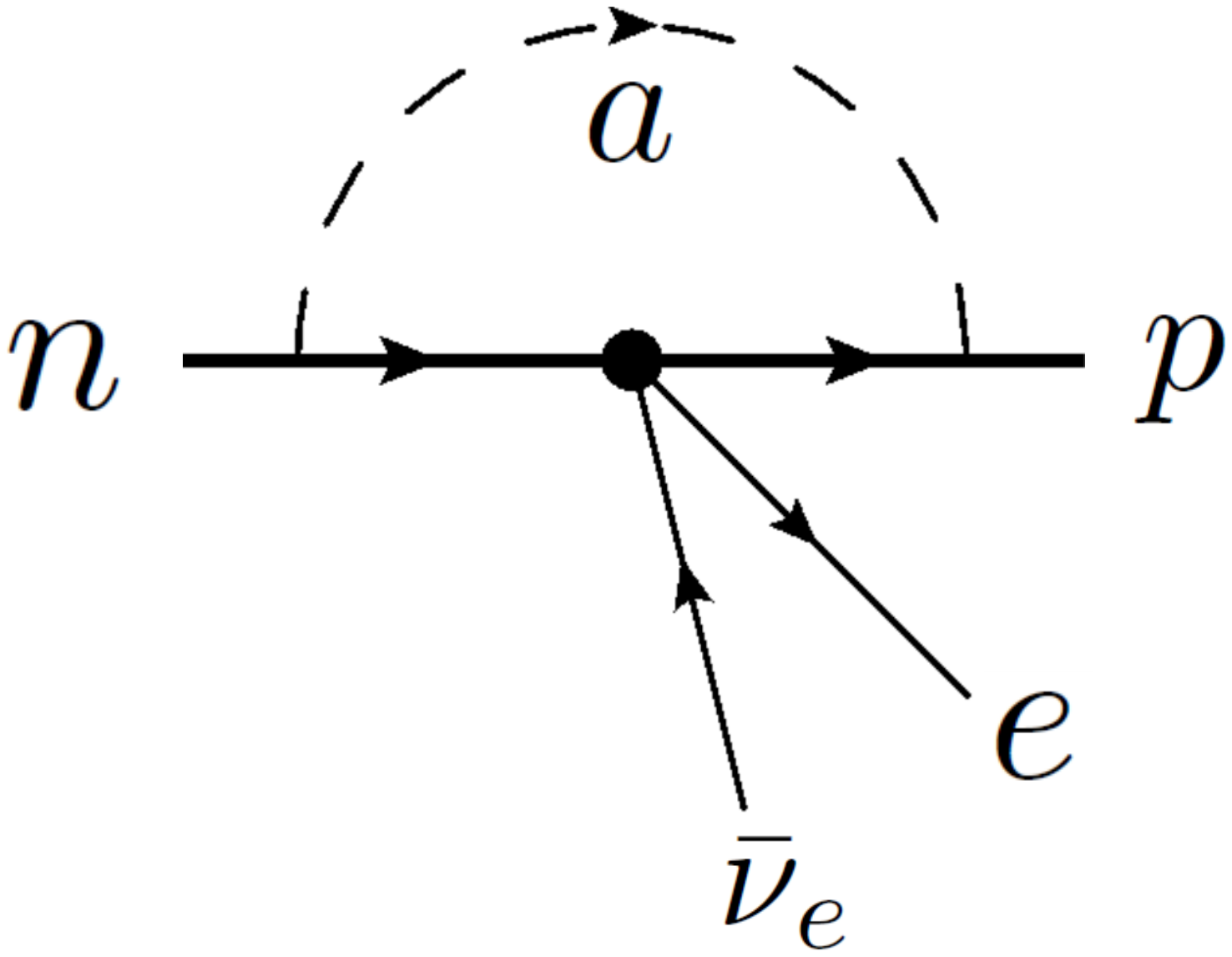}
		\par\end{centering}
	\caption{\label{fig:axion} The axion-induced one-particle-irreducible diagram.}
\end{figure}

In Table~\ref{tab:zerosum} we display a list of correlation coefficient pairs that sum up to zero in the EFT description; here we provide a simple example to show how the inclusion of a light DOF can break this ``zero-sum game'' which provides experimental signatures of its existence. Notice that this is just for illustrative purpose and we do not intend to perform a more comprehensive analysis that may have to account for SM corrections and real constraints from other low-energy experiments. 

We consider an axion field $\boldsymbol{a}$ with mass $m_{\boldsymbol{a}}$ that couples to the nucleons as:
\begin{equation}
	\mathcal{L}_{\boldsymbol{a}N}=-i\boldsymbol{a}(g_p\bar{p}\gamma_5 p+g_n\bar{n}\gamma_5 n)~,
\end{equation}
where $g_p$, $g_n$ are dimensionless coupling constants. Assuming $m_{\boldsymbol{a}}>m_n-m_p$, it cannot be emitted as a real particle in the neutron beta decay and can only appear in loops. The axion-induced nucleon wavefunction renormalization provides only a constant multiplicative correction to $\xi$ which is irrelevant to our discussion of correlation coefficients (since $\xi$ is always divided out), so we may focus on the one-particle-irreducible diagram depicted in Fig.\ref{fig:axion}. It gives the following correction to the decay amplitude:
\begin{equation}
	\delta\mathcal{M}(p_p,p_n)=-i\frac{G_V}{\sqrt{2}}g_pg_n L_\lambda \mu^{4-d}\int\frac{d^dk}{(2\pi)^d}\frac{\bar{u}_p\slashed{k}\gamma^\lambda(1+\lambda \gamma_5)\slashed{k}u_n}{((p_p-k)^2-m_p^2)((p_n-k)^2-m_n^2)(k^2-m_{\boldsymbol{a}}^2)}~,\label{eq:deltaM}
\end{equation}
where $L_\lambda$ is the matrix element of the leptonic weak current, and we have regulated the ultraviolet (UV) divergence using dimensional regularization, and simplified the numerator of the integrand using the equation of motion. It is easy to see that, the UV-divergent part of the integral serves only to renormalize the SM couplings $G_V$ and $\lambda$. In fact, since these couplings are usually defined in terms of the forward nucleon matrix element, one may in stead work on the following UV-finite renormalized amplitude:
\begin{equation}
\delta\mathcal{M}_r(p_p,p_n)\equiv \delta \mathcal{M}(p_p,p_n)-\delta\mathcal{M}(p,p)~,
\end{equation}
by reabsorbing the forward piece $\delta \mathcal{M}(p,p)$ into the definition of $G_V$ and $\lambda$. 

\begin{figure}
	\begin{centering}
		\includegraphics[scale=0.4]{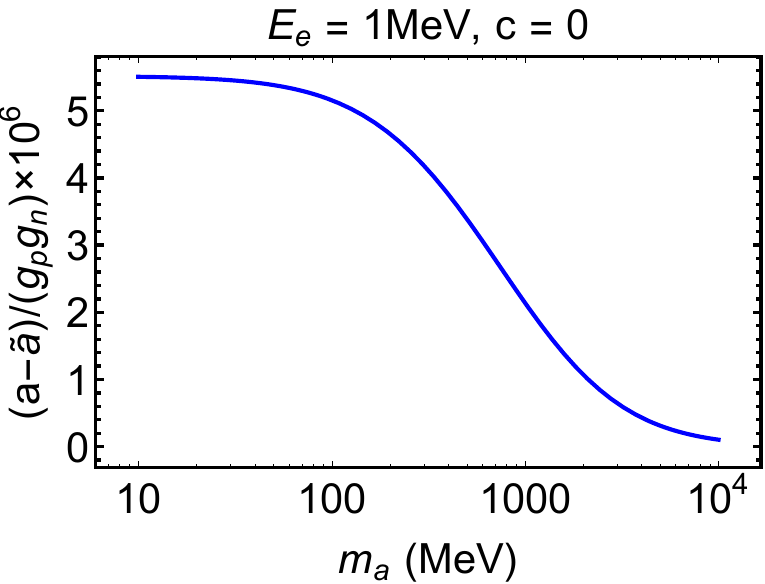}
		\includegraphics[scale=0.4]{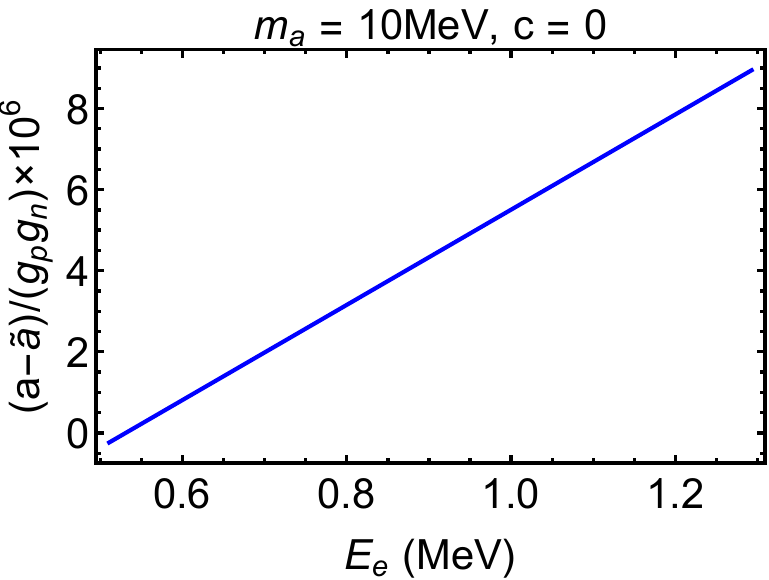}
		\includegraphics[scale=0.4]{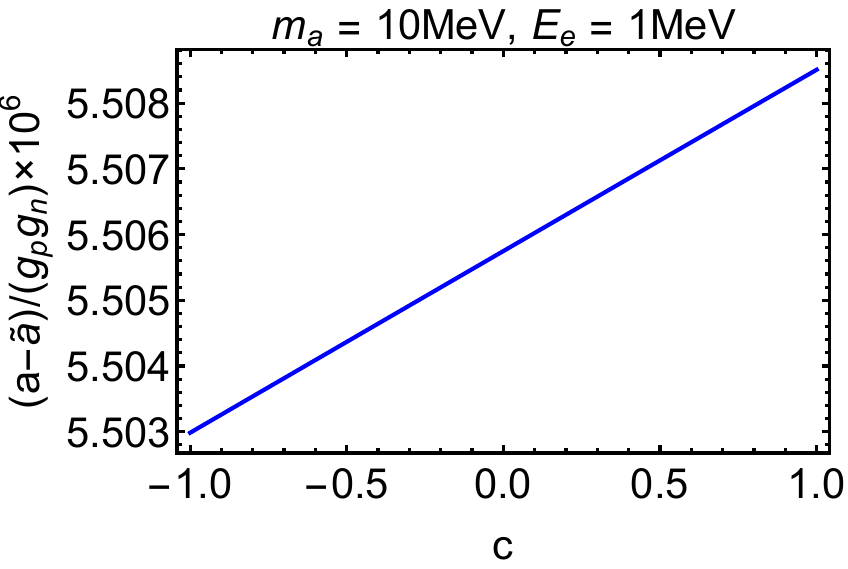}
	\par\end{centering}
	\caption{\label{fig:breaking} The axion-induced breaking of the $a=\tilde{a}$ equality as functions of $m_{\boldsymbol{a}}$, $E_e$ and $c$.}
\end{figure}

We recall that, in a pure LY-induced beta decay (after subtracting the known higher-order SM corrections) the correlation coefficient $a$ is exactly equal to $\tilde{a}$; the former is a well-studied correlation and the latter is a new correlation introduced in this paper.  
However, the inclusion of the correction $\delta\mathcal{M}_r$
results in a finite breaking of such an equality. In Fig.\ref{fig:breaking} we plot the results of the difference $a-\tilde{a}$ with the axion coupling constants $g_p g_n$ scaled out. Notice that now the correlation coefficients are no longer constants but functions of $E_e$ and $c\equiv \hat{p}_e\cdot\hat{p}_\nu$, as well as the axion mass $m_{\boldsymbol{a}}$, so we show three plots to illustrate their functional dependence on all the three variables.

\section{Summary}

To summarize, we studied the SM prediction and the leading BSM correction, within the EFT framework, of the free neutron differential decay rate assuming that the proton polarization is measurable. We showed that it gives rise to a very rich phenomenology that can be used for the precision test of the SM, the search for new physics and the test of the EFT framework. This may provide new motivations for future experimental designs to measure the proton polarization. A generalization to allowed nuclear beta decays is also possible but will not be attempted in this work.
	
\begin{acknowledgments}
		
We thank Hartmut Abele for inspiring discussions, and are grateful to Alejandro Garcia and Vincenzo Cirigliano for carefully reading the manuscript and providing invaluable suggestions. The work of C.-Y.S. is supported in
part by the U.S. Department of Energy (DOE), Office of Science, Office of Nuclear Physics, under the FRIB Theory Alliance award DE-SC0013617, and by the DOE grant DE-FG02-97ER41014. We acknowledge support from the DOE Topical Collaboration ``Nuclear Theory for New Physics'', award No. DE-SC0023663. 
		
\end{acknowledgments}
	
\begin{appendix}
		
\section{\label{sec:zeroth}Pure Lee-Yang correlation coefficients}

In this Appendix we provide the analytic formula for the purely LY-induced correlation coefficients. Those from $\xi$ and  $g_{\text{JTW}}$ read:
\begingroup
\allowdisplaybreaks
\begin{eqnarray}
	(\xi)_\text{LY}&=&|C_S|^2+|C_V|^2+|C_S'|^2+|C_V'|^2+3\left(|C_T|^2+|C_A|^2+|C_T'|^2+|C_A'|^2\right)\nonumber\\
	(a\xi)_\text{LY}&=&-|C_S|^2+|C_V|^2-|C_S'|^2+|C_V'|^2+|C_T|^2-|C_A|^2+|C_T'|^2-|C_A'|^2\nonumber\\
	(b\xi)_\text{LY}&=&2\mathfrak{Re}\left\{C_SC_V^*+C_S'C_V^{\prime *}+3\left(C_T C_A^*+C_T'C_A^{\prime *}\right)\right\}\nonumber\\
	(A\xi)_\text{LY}&=&2\mathfrak{Re}\left\{2\left(C_TC_T^{\prime *}-C_AC_A^{\prime *}\right)+C_SC_T^{\prime *}+C_S'C_T^*-C_VC_A^{\prime *}-C_V'C_A^*\right\}\nonumber\\
	(B\xi)_\text{LY}&=&2\mathfrak{Re}\left\{2\left[\frac{m_e}{E_e}\left(C_TC_A^{\prime *}+C_T'C_A^*\right)+\left(C_TC_T^{\prime *}+C_AC_A^{\prime *}\right)\right]-C_SC_T^{\prime *}-C_S'C_T^*\right.\nonumber\\
	&&\left.-C_VC_A^{\prime *}-C_V'C_A^*-\frac{m_e}{E_e}\left(C_SC_A^{\prime *}+C_S'C_A^*+C_VC_T^{\prime *}+C_V'C_T^*\right)\right\}\nonumber\\
	(D\xi)_\text{LY}&=&2\mathfrak{Im}\left\{C_SC_T^*-C_VC_A^*+C_S'C_T^{\prime *}-C_V'C_A^{\prime *}\right\}\nonumber\\
	(G\xi)_\text{LY}&=&2\mathfrak{Re}\left\{C_SC_S^{\prime *}-C_VC_V^{\prime *}+3\left(C_TC_T^{\prime *}-C_AC_A^{\prime *}\right)\right\}\nonumber\\
	(H\xi)_\text{LY}&=&2\mathfrak{Re}\left\{-C_SC_V^{\prime *}-C_S'C_V^*+C_TC_A^{\prime *}+C_T'C_A^*+\frac{m_e}{E_e}\left(C_TC_T^{\prime *}+C_AC_A^{\prime*}-C_SC_S^{\prime *}-C_VC_V^{\prime*}\right)\right\}\nonumber\\
	(K\xi)_\text{LY}&=&2\mathfrak{Re}\left\{-C_SC_S^{\prime*}-C_VC_V^{\prime *}+C_SC_V^{\prime *}+C_S'C_V^*+C_TC_T^{\prime*}+C_AC_A^{\prime *}-C_TC_A^{\prime *}-C_T'C_A^*\right\}\nonumber\\
	(L\xi)_\text{LY}&=&2\mathfrak{Im}\left\{C_SC_V^*+C_S'C_V^{\prime *}-C_TC_A^*-C_T'C_A^{\prime *}\right\}\nonumber\\
	(N\xi)_\text{LY}&=&2\mathfrak{Re}\left\{\frac{m_e}{E_e}\left(|C_T|^2+|C_T'|^2+|C_A|^2+|C_A'|^2\right)+2\left(C_TC_A^*+C_T'C_A^{\prime *}\right)\right.\nonumber\\
	&&\left.+C_SC_A^*+C_VC_T^*+C_S'C_A^{\prime *}+C_V'C_T^{\prime *}+\frac{m_e}{E_e}\left(C_SC_T^*+C_VC_A^*+C_S'C_T^{\prime *}+C_V'C_A^{\prime *}\right)\right\}\nonumber\\
	(Q\xi)_\text{LY}&=&2\mathfrak{Re}\left\{|C_T|^2+|C_A|^2+|C_T'|^2+|C_A'|^2-2\left(C_TC_A^*+C_T'C_A^{\prime *}\right)\right.\nonumber\\
	&&\left.-C_SC_A^*-C_VC_T^*-C_S'C_A^{\prime *}-C_V'C_T^{\prime *}+C_SC_T^*+C_VC_A^*+C_S'C_T^{\prime *}+C_V'C_A^{\prime *}\right\}\nonumber\\
	(R\xi)_\text{LY}&=&2\mathfrak{Im}\left\{2\left(C_TC_A^{\prime *}+C_T'C_A^*\right)+C_SC_A^{\prime *}+C_S'C_A^*-C_VC_T^{\prime *}-C_V'C_T^*\right\}
\end{eqnarray}
\endgroup
Those from $g_\text{EF}$ read:
\begingroup
\allowdisplaybreaks
\begin{eqnarray}
	(S\xi)_\text{LY}&=&2\mathfrak{Re}\left\{-C_SC_A^*-C_S'C_A^{\prime *}+C_VC_T^*+C_V'C_T^{\prime *}\right\}\nonumber\\
	(T\xi)_\text{LY}&=&2\mathfrak{Re}\left\{-C_SC_T^*-C_S'C_T^{\prime *}+C_VC_A^*+C_V'C_A^{\prime *}+|C_T|^2+|C_T'|^2-|C_A|^2-|C_A'|^2\right\}\nonumber\\
	(U\xi)_\text{LY}&=&2\mathfrak{Re}\left\{C_SC_A^*+C_S'C_A^{\prime *}-C_VC_T^*-C_V'C_T^{\prime *}\right\}\nonumber\\
	(V\xi)_\text{LY}&=&2\mathfrak{Im}\left\{\frac{m_e}{E_e}\left(C_SC_T^{\prime *}+C_S'C_T^{\prime *}+C_VC_A^{\prime *}+C_V'C_A^*\right)+C_SC_A^{\prime *}+C_S'C_A^*+C_VC_T^{\prime *}+C_V'C_T^*\right\}\nonumber\\
	(W\xi)_\text{LY}&=&2\mathfrak{Im}\left\{C_SC_T^{\prime *}+C_S'C_T^*+C_VC_A^{\prime *}+C_V'C_A^*-C_SC_A^{\prime *}-C_S'C_A^*-C_VC_T^{\prime *}-C_V'C_T^*\right\}
\end{eqnarray}
\endgroup
Those from $g_{s_p}$ read:
\begingroup
\allowdisplaybreaks
\begin{eqnarray}
	(\tilde{A}\xi)_\text{LY}&=&2\mathfrak{Re}\left\{-2\left(C_TC_T^{\prime *}-C_AC_A^{\prime *}\right)+C_SC_T^{\prime *}+C_S'C_T^*-C_VC_A^{\prime *}-C_V'C_A^*\right\}\nonumber\\
	(\tilde{B}\xi)_\text{LY}&=&-2\mathfrak{Re}\left\{2\left[\frac{m_e}{E_e}\left(C_TC_A^{\prime *}+C_T'C_A^*\right)+\left(C_TC_T^{\prime *}+C_AC_A^{\prime *}\right)\right]+C_SC_T^{\prime *}+C_S'C_T^*\right.\nonumber\\
	&&\left.+C_VC_A^{\prime *}+C_V'C_A^*+\frac{m_e}{E_e}\left(C_SC_A^{\prime *}+C_S'C_A^*+C_VC_T^{\prime *}+C_V'C_T^*\right)\right\}\nonumber\\
	(\tilde{D}\xi)_\text{LY}&=&(D\xi)_\text{LY}\nonumber\\
	(\tilde{N}\xi)_\text{LY}&=&2\mathfrak{Re}\left\{-\frac{m_e}{E_e}\left(|C_T|^2+|C_T'|^2+|C_A|^2+|C_A'|^2\right)-2\left(C_TC_A^*+C_T'C_A^{\prime *}\right)\right.\nonumber\\
	&&\left.+C_SC_A^*+C_VC_T^*+C_S'C_A^{\prime *}+C_V'C_T^{\prime *}+\frac{m_e}{E_e}\left(C_SC_T^*+C_VC_A^*+C_S'C_T^{\prime *}+C_V'C_A^{\prime *}\right)\right\}\nonumber\\
	(\tilde{Q}\xi)_\text{LY}&=&2\mathfrak{Re}\left\{-|C_T|^2-|C_A|^2-|C_T'|^2-|C_A'|^2+2\left(C_TC_A^*+C_T'C_A^{\prime *}\right)\right.\nonumber\\
	&&\left.-C_SC_A^*-C_VC_T^*-C_S'C_A^{\prime *}-C_V'C_T^{\prime *}+C_SC_T^*+C_VC_A^*+C_S'C_T^{\prime *}+C_V'C_A^{\prime *}\right\}\nonumber\\
	(\tilde{R}\xi)_\text{LY}&=&2\mathfrak{Im}\left\{-2\left(C_TC_A^{\prime *}+C_T'C_A^*\right)+C_SC_A^{\prime *}+C_S'C_A^*-C_VC_T^{\prime *}-C_V'C_T^*\right\}\nonumber\\
	(\tilde{S}\xi)_\text{LY}&=&(S\xi)_\text{LY}\nonumber\\
	(\tilde{T}\xi)_\text{LY}&=&2\mathfrak{Re}\left\{-C_SC_T^*-C_S'C_T^{\prime *}+C_VC_A^*+C_V'C_A^{\prime *}-|C_T|^2-|C_T'|^2+|C_A|^2+|C_A'|^2\right\}\nonumber\\
	(\tilde{U}\xi)_\text{LY}&=&(U\xi)_\text{LY}\nonumber\\
	(\tilde{V}\xi)_\text{LY}&=&(V\xi)_\text{LY}\nonumber\\
	(\tilde{W}\xi)_\text{LY}&=&(W\xi)_\text{LY}
\end{eqnarray}
\endgroup
And finally, those from $g_{s_ps_n}$ read:
\begingroup
\allowdisplaybreaks
\begin{eqnarray}
	(X\xi)_\text{LY}&=&|C_S|^2+|C_S'|^2+|C_V|^2+|C_V'|^2-|C_T|^2-|C_T'|^2-|C_A|^2-|C_A'|^2\nonumber\\
	(\tilde{a}\xi)_\text{LY}&=&-|C_S|^2-|C_S'|^2+|C_V|^2+|C_V'|^2+|C_T|^2+|C_T'|^2-|C_A|^2-|C_A'|^2\nonumber\\
	(\tilde{b}\xi)_\text{LY}&=&2\mathfrak{Re}\left\{C_SC_V^*+C_S'C_V^{\prime *}-C_TC_A^*-C_T'C_A^{\prime *}\right\}\nonumber\\
	(\tilde{G}\xi)_\text{LY}&=&2\mathfrak{Re}\left\{C_SC_S^{\prime *}-C_TC_T^{\prime *}-C_VC_V^{\prime *}+C_AC_A^{\prime *}\right\}\nonumber\\
	(\tilde{H}\xi)_\text{LY}&=&2\mathfrak{Re}\left\{-C_SC_V^{\prime *}-C_S'C_V^*+C_TC_A^{\prime *}+C_T'C_A^*+\frac{m_e}{E_e}\left[-C_SC_S^{\prime *}+C_TC_T^{\prime *}-C_VC_V^{\prime *}+C_AC_A^{\prime Ü}\right]\right\}\nonumber\\
	(\tilde{K}\xi)_\text{LY}&=&2\mathfrak{Re}\left\{-C_SC_S^{\prime *}-C_VC_V^{\prime *}+C_TC_T^{\prime *}+C_AC_A^{\prime *}+C_SC_V^{\prime *}+C_S'C_V^*-C_TC_A^{\prime *}-C_T'C_A^*\right\}\nonumber\\
	(\tilde{L}\xi)_\text{LY}&=&2\mathfrak{Im}\left\{C_SC_V^*+C_S'C_V^{\prime *}-C_TC_A^*-C_T'C_A^{\prime *}\right\}\nonumber\\
	(Y\xi)_\text{LY}&=&-2\mathfrak{Im}\left\{C_SC_A^*+C_S'C_A^{\prime *}+C_VC_T^*+C_V'C_T^{\prime *}+\frac{m_e}{E_e}\left[C_SC_T^*+C_S'C_T^{\prime *}+C_VC_A^*+C_V'C_A^{\prime*}\right]\right\}\nonumber\\
	(\text{\k{A}}\xi)_\text{LY}&=&2\mathfrak{Re}\bigg\{C_SC_A^{\prime*}+C_S'C_A^*+C_VC_T^{\prime *}+C_V'C_T^*-2\left(C_TC_A^{\prime *}+C_T'C_A^*\right)\nonumber\\
	&&+\frac{m_e}{E_e}\left[C_SC_T^{\prime*}+C_S'C_T^*+C_VC_A^{\prime *}+C_V'C_A^*-2\left(C_TC_T^{\prime*}+C_AC_A^{\prime*}\right)\right]\bigg\}\nonumber\\
	(\text{\k{a}}\xi)_\text{LY}&=&-2\mathfrak{Re}\bigg\{C_SC_A^{\prime*}+C_S'C_A^*+C_VC_T^{\prime *}+C_V'C_T^*+2\left(C_TC_A^{\prime *}+C_T'C_A^*\right)\nonumber\\
	&&+\frac{m_e}{E_e}\left[C_SC_T^{\prime*}+C_S'C_T^*+C_VC_A^{\prime *}+C_V'C_A^*+2\left(C_TC_T^{\prime*}+C_AC_A^{\prime*}\right)\right]\bigg\}\nonumber\\
	(\text{\k{E}}\xi)_\text{LY}&=&2\mathfrak{Re}\left\{-C_SC_A^{\prime*}-C_S'C_A^*+C_VC_T^{\prime*}+C_V'C_T^*\right\}\nonumber\\
	(\text{\k{e}}\xi)_\text{LY}&=&2\mathfrak{Re}\left\{C_SC_A^{\prime*}+C_S'C_A^*-C_VC_T^{\prime*}-C_V'C_T^*\right\}\nonumber\\
	(\text{\L{}}\xi)_\text{LY}&=&2\mathfrak{Im}\left\{C_SC_T^{\prime*}+C_S'C_T^*+C_VC_A^{\prime*}+C_V'C_A^*+\frac{m_e}{E_e}\left[C_SC_A^{\prime*}+C_S'C_A^*+C_VC_T^{\prime*}+C_V'C_T^*\right]\right\}\nonumber\\
	(\text{\l{}}\xi)_\text{LY}&=&2\mathfrak{Im}\left\{-C_SC_T^{\prime*}-C_S'C_T^*+C_VC_A^{\prime*}+C_V'C_A^*\right\}\nonumber\\
	(\text{\'{N}}\xi)_\text{LY}&=&2\mathfrak{Re}\left\{-|C_T|^2-|C_T'|^2+|C_A|^2+|C_A'|^2-C_SC_T^*-C_S'C_T^{\prime*}+C_VC_A^*+C_V'C_A^{\prime*}\right\}\nonumber\\
	(\text{\'{n}}\xi)_\text{LY}&=&2\mathfrak{Re}\left\{-|C_T|^2-|C_T'|^2+|C_A|^2+|C_A'|^2+C_SC_T^*+C_S'C_T^{\prime*}-C_VC_A^*-C_V'C_A^{\prime*}\right\}\nonumber\\
	(\text{\'{O}}\xi)_\text{LY}&=&2\mathfrak{Im}\left\{C_SC_A^*+C_S'C_A^{\prime*}-C_VC_T^*-C_V'C_T^{\prime*}+2\left(C_TC_A^*+C_T'C_A^{\prime*}\right)\right\}\nonumber\\
	(\text{\'{o}}\xi)_\text{LY}&=&2\mathfrak{Im}\left\{-C_SC_A^*-C_S'C_A^{\prime*}+C_VC_T^*+C_V'C_T^{\prime*}+2\left(C_TC_A^*+C_T'C_A^{\prime*}\right)\right\}\nonumber\\
	(\text{\'{S}}\xi)_\text{LY}&=&2\mathfrak{Im}\left\{C_SC_T^*+C_S'C_T^{\prime*}-C_VC_A^*-C_V'C_A^{\prime*}\right\}\nonumber\\
	(\text{\'{s}}\xi)_\text{LY}&=&2\mathfrak{Im}\left\{-C_SC_T^*-C_S'C_T^{\prime*}+C_SC_A^*+C_S'C_A^{\prime*}+C_VC_T^*+C_V'C_T^{\prime*}-C_VC_A^*-C_V'C_A^{\prime*}\right\}\nonumber\\
	(\text{\'{Z}}\xi)_\text{LY}&=&2\mathfrak{Re}\left\{-C_SC_T^{\prime*}-C_S'C_T^*+C_SC_A^{\prime*}+C_S'C_A^*+C_VC_T^{\prime*}+C_V'C_T^*-C_VC_A^{\prime*}-C_V'C_A^*\right.\nonumber\\
	&&\left.+2\left(-C_TC_T^{\prime*}-C_AC_A^{\prime*}+C_TC_A^{\prime*}+C_T'C_A^*\right)\right\}\nonumber\\
	(\text{\'{z}}\xi)_\text{LY}&=&2\mathfrak{Re}\left\{C_SC_T^{\prime*}+C_S'C_T^*-C_SC_A^{\prime*}-C_S'C_A^*-C_VC_T^{\prime*}-C_V'C_T^*+C_VC_A^{\prime*}+C_V'C_A^*\right.\nonumber\\
	&&\left.+2\left(-C_TC_T^{\prime*}-C_AC_A^{\prime*}+C_TC_A^{\prime*}+C_T'C_A^*\right)\right\}
\end{eqnarray}
\endgroup

\end{appendix}

\bibliography{ref}

\end{document}